\documentclass[12pt,preprint,letterpaper]{emulateapj}

\def\nthp {N$_2$H$^+$}
\def\SINGLESEDBETA {2.3}
\def\SINGLESEDBETARMS {0.4}
\def\SINGLESEDTEMP {8.9}
\def\SINGLESEDTEMPRMS {1.0}
\def\SINGLESEDNH {10.2$\times$10$^{21}$}
\def\SINGLESEDNHRMS {2.1$\times$10$^{21}$}
\def\MAPSINGLEBETA {2.2}
\def\MAPSINGLEBETARMS {0.6}
\def\MAPMULTIBETA {2.0}
\def\MAPMULTIBETARMS {0.6}
\def\MAPMULTITEMP {9.4}
\def\MAPMULTITEMPRMS {1.9}
\def\MODELBETA {2.0}
\def\MODELBETARMS {0.4}

\shorttitle{Emissivity Spectral Index in TMC-1C}
\shortauthors{Schnee et al.}

\begin{document}

\title{The Dust Emissivity Spectral Index in the Starless Core TMC-1C} 

\author{Scott Schnee\altaffilmark{1,8}, Melissa Enoch\altaffilmark{2},
Alberto Noriega-Crespo\altaffilmark{3}, Jack Sayers\altaffilmark{4},
Susan Terebey\altaffilmark{5}, Paola Caselli\altaffilmark{6}, Jonathan
Foster\altaffilmark{7}, Alyssa Goodman\altaffilmark{7}, Jens
Kauffmann\altaffilmark{7}, Deborah Padgett\altaffilmark{3}, Luisa
Rebull\altaffilmark{3}, Anneila Sargent\altaffilmark{1}, Rahul
Shetty\altaffilmark{7}}

\email{scott.schnee@nrc-cnrc.gc.ca}

\altaffiltext{1}{Department of Astronomy, California Institute of
Technology, MC 105-24 Pasadena, CA 91125, USA}
\altaffiltext{2}{Department of Astronomy, University of California,
Berkeley, CA 94720, USA}
\altaffiltext{3}{{\it Spitzer} Science Center, MC 220-6, California
Institute of Technology, Pasadena, CA 91125, USA}
\altaffiltext{4}{Jet Propulsion Laboratory, California Institute of
Technology, 4800 Oak Grove Drive, Pasadena, CA 9110, USA9}
\altaffiltext{5}{Department of Physics and Astronomy PS315, 5151 State
University Drive, California State University at Los Angeles, Los
Angeles, CA 90032, USA}
\altaffiltext{6}{School of Physics and Astronomy, University of Leeds,
Leeds LS2 9JT}
\altaffiltext{7}{Harvard-Smithsonian Center for Astrophysics, 60
Garden Street, Cambridge, MA 02138, USA}
\altaffiltext{8}{Current address: NRC-HIA, 5071 West Saanich Road
Victoria, BC V9E 2E7, Canada}

\begin{abstract}

In this paper we present a dust emission map of the starless core
TMC-1C taken at 2100~\micron.  Along with maps at 160, 450, 850 and
1200~\micron, we study the dust emissivity spectral index from the
(sub)millimeter spectral energy distribution, and find that it is
close to the typically assumed value of $\beta = 2$.  We also map the
dust temperature and column density in TMC-1C, and find that at the
position of the dust peak ($A_V \sim 50$), the line-of-sight-averaged
temperature is $\sim$7~K.  Employing simple Monte Carlo modeling, we
show that the data are consistent with a constant value for the
emissivity spectral index over the whole map of TMC-1C.

\end{abstract}


\section{Introduction}

Starless cores are often identified through maps of their dust
emission at (sub)millimeter wavelengths, and recent surveys of nearby
molecular clouds have increased the number of such objects
considerably \citep[e.g.][]{Hatchell07, Enoch08, Kauffmann08,
Simpson08}.  Analysis of the properties of starless cores is often
restricted by sparse sampling of their spectral energy distribution
(SED), uncertain dust emissivities and assumptions about their
geometries.  As a result, the measured (sub)millimeter fluxes are
often used in conjunction with assumed values for the dust temperature
and/or emissivity to determine core masses, though the systematic
errors can be substantial (a factor of a few).  With sufficient
spectral coverage and knowlege of source geometry, it should be
possible to map the column density, temperature and dust emissivity
spectral index in a starless core, though this has yet to be
accomplished.

One of the best-studied starless cores is TMC-1C, located in the
Taurus molecular cloud at an approximate distance of 140 pc
\citep{Torres09}.  Previous studies of the dust emission from TMC-1C
at 450, 850 and 1200 \micron\ have determined that it is cold ($T_d
\sim$6 K) and dense ($n \sim$10$^6$ cm$^{-3}$) at its center, and
becomes progressively less dense and warmer at larger radii
\citep{Schnee07a, Schnee05}.  The dust-derived mass of TMC-1C is a
factor of a few larger than its virial mass, and self-absorbed \nthp\
spectra show evidence for sub-sonic infall \citep{Schnee07b,
Schnee05}.  The chemistry of TMC-1C is characterized by depletion of
CO and its isotopologues, as well as depletion of CS and \nthp\
\citep{Schnee07b}.

Previously reported mass and temperature profiles for TMC-1C are
estimated assuming a constant value for the emissivity spectral index
($\beta = 1.8$) \citep{Schnee07a}.  However, a wide range of values
for $\beta$ in other regions have been measured or calculated on
theoretical grounds.  For instance, in the diffuse interstellar medium
(ISM), an emissivity spectral index of $\beta = 2$ fits the typical
far-infrared spectrum well \citep{Draine07, Draine84}, while grain
growth in protostellar disks results in values of $\beta \simeq 1$
\citep{Beckwith91}.  Since starless cores are in an intermediate stage
between the diffuse ISM and protostars, one might expect that the dust
in a starless core would have an emissivity spectral index in the
range $1 \le \beta \le 2$, although the timescale of grain growth at
densities $n_H \sim 10^6$ cm$^{-3}$ is several hundred times the
free-fall time of a typical starless core \citep{Chakrabarti05}.  On
the other hand, the growth of icy mantles on dust grains could steepen
the slope of the dust SED, and this has been cited as the reason for
high values of the emissivity spectral index in parts of the Orion
Ridge ($\beta \sim 2.5$) \citep{Lis98} and in Sgr B2(N) ($\beta \sim
3.7$) \citep{Kuan96}.  In the cold and dense evironment in TMC-1C, the
dust grains could form similar icy mantles, which might drive the
emissivity spectral index to values $\beta > 2$.

In addition to changes in $\beta$, grain growth by either coagulation
or accretion of an icy mantle will change the opacity of the dust
grains, $\kappa$.  For instance, evidence that dust opacities are
higher in dense cores than in the diffuse ISM can be found by
comparing gas temperature and dust emission profiles \citep{Keto08} or
by comparing dust and gas emission profiles \citep{Keto04}.  From
models of dust coagulation for conditions that are plausible for a
starless core, \citet{Ossenkopf94} show that the dust opacity at
1.3~mm can be a factor of five higher than in a diffuse environment.
Using 160~\micron\ data for Taurus, \citet{Terebey09} measure an
opacity of $\kappa_{160~\micron} = 0.23 \pm 0.046$ cm$^2$g$^{-1}$ in
the cold cloud ($A_V$ ranging from 0.4 to 4.0).  The opacity is 2.6
times greater than the diffuse ISM opacity \citep{Weingartner03}, and
provides evidence for higher opacity values in dense regions
\citep[c.f.][]{Ossenkopf94}.

Laboratory studies of possible dust grain materials have measured the
dust emission and absorption properties at far-infrared and longer
wavelengths.  For instance, \citet{Aannestad75} measured the
absorption of silicate grains and found that $\beta \simeq 2$ for
grains without an ice mantle, and $\beta \simeq 3$ for grains with an
ice mantle.  The steepest spectral indices calculated by
\citet{Aannestad75} were for fused quartz and olivine with ice
mantles, both of which had $\beta = 3.5$, close to the extremely steep
$\beta = 3.7$ reported by \citet{Kuan96}.  In a study of the
millimeter-wave absorption of silicate grains in the temperature range
1.2 - 30~K, \citet{Agladze96} found that the emissivity spectral index
depends on the dust temperature and for some materials can vary
between $1.5 \le \beta \le 2.5$, with the largest value for $\beta$ at
a temperature $T_d \sim 10$~K.  A temperature-dependent emissivity
spectral index was also found by \citet{Mennella98} for cosmic dust
analog grains in the temperature range $24 \le T_d \le 295$~K and a
wavelength range $20~\micron \le \lambda \le 2$~mm, with a steeper SED
at lower temperatures.  The range of emissivity spectral indices is
$1.2 \le \beta \le 2.4$ for the various grain types considered by
\citet{Mennella98}.  In a study of amorphous silicates in the
temperature range $10 \le T_d \le 300$~K and wavelength range $0.1 \le
\lambda \le 2$~mm, \citet{Boudet05} also report an anticorrelation
between $T_d$ and $\beta$, with values of the emissivity spectral
index between $1.5 \le \beta \le 2.5$.  Based on these laboratory
measurements, one might expect that in starless cores like TMC-1C,
which have temperatures close to 10~K \citep{Schnee09}, the emissivity
spectral index would be close to $\beta \simeq 2.5$.

Observations of nearby star-forming regions suggest an anticorrelation
between $T_d$ and $\beta$ similar to that measured in the laboratory.
For instance, observations of the Orion, $\rho$ Ophiuchi and Taurus
molecular clouds at far-infrared and (sub)millimeter wavelengths by
\citet{Dupac03} show that the emissivity spectral index may vary from
$2.5 \ge \beta \ge 1.0$ as the dust temperature rises from $10 \le T_d
\le 80$~K.  Similarly, in a study of the far-infrared and
(sub)millimeter SEDs of luminous infrared galaxies, \citet{Yang07}
report that the emissivity spectral index varies from $2.4 \ge \beta
\ge 0.9$ as the dust temperature varies from $30 \le T_d \le 60$~K.
In a study of the starless core L1498, \citet{Shirley05} find an
emissivity spectral index of $\beta = 2.44 \pm 0.62$ for an assumed
dust temperature of $T_d = 10.5 \pm 0.5$~K, consistent with the
hypothesis that the cold dust in a starless core has a steep spectral
index.

However, variations in the dust properties along the line-of-sight and
noise in the observed maps can mimic an anticorrelation between $T_d$
and $\beta$ \citep{Shetty09a, Shetty09b}.  Furthermore, at high
densities and low temperatures, grain coagulation will decrease
$\beta$ while the possible $T_d$--$\beta$ relation will increase the
emissivity spectral index.  Therefore, it is not at all clear that one
should expect the emissivity spectral index to vary significantly in a
starless core.  Observations of dust emission and extinction in the
dark cloud B68 show no evidence for variations in the dust properties
\citep{Bianchi03}, suggesting that to some extent the various
pressures on $\beta$ may cancel out.

In this paper we present maps of the starless core TMC-1C at 160, 450,
850, 1200 and 2100~\micron.  By spatially resolving the emission from
TMC-1C over a wide range of wavelengths, we then map the column
density, temperature and emissivity spectral index of the dust, which
has typically been difficult to do with previous datasets.

\section{Observations} \label{OBSERVATIONS}

The 160, 450, 850 and 1200~\micron\ maps have been presented or
described in previous papers \citep{Terebey09, Schnee05, Schnee07a,
diFrancesco08, Kauffmann08}, so here we briefly review the
observations and data reduction, and refer the reader to the earlier
papers for more details.  The 2100~\micron\ map has not been
previously presented, so it is discussed in detail below.

\subsection{Spitzer}

The Spitzer data near TMC-1C derive from the Taurus Spitzer Legacy
Survey \citep{Padgett09}, a large $44$ square degree map of the Taurus
star-forming region at seven wavelengths ranging from 3.6 to
160~\micron.  See \citet{Padgett09} for an overview of the survey
parameters and data processing. The MIPS data utilize the fast scan
mode with 3 seconds integration time per frame and 5 visits per
pixel. The MIPS 160~\micron\ data obtained in fast scan mode mapping
does not have enough redundancy to fill in completely all the data
gaps due to a dead readout and other effects. To mitigate these data
gaps and to preserve the diffuse emission as much as possible, a 3
pixel by 3 pixel (15\arcsec\ per pixel) median filter is applied to
the image. The median filter introduces some image smoothing,
increasing the effective angular resolution from 40\arcsec\ to
1\arcmin.  In some cases, filtering reduces the surface brightness by
10--15\%, though the reduction in the map of TMC-1C is only 2\%.

Analysis of the 160~\micron\ data in Taurus shows there are two main
cloud components: emission from the extended cold cloud ($14.2 \pm
0.4$~K , $A_V < 3$), and emission from compact cold cores ($T_d
\sim$10~ K, high $A_V$) \citep{Terebey09, Flagey09}. Since our present
focus is the high extinction regions sampled by (chopped) millimeter
data, here we further process the 160~\micron\ data to remove the
spatially extended cloud emission. Following the procedure in
\citet{Terebey09} we use IRAS 100~\micron\ images to model and
subtract the 14.2~K cloud emission. The resulting 160~\micron\ image
traces the high extinction ridge and looks very similar to the
millimeter continuum emission.  The median flux removed from original
160~\micron\ map is 96 MJy/sr, or about 90\% of the total flux.
Uncertainties in the temperature of the extended cloud component
results in a 10\% uncertainty in the final 160~\micron\ map.  Given
the 12\% absolute flux uncertainty of MIPS 160~\micron\ maps
\citep{Stansberry07}, we conservatively estimate the absolute
uncertainty in the 160~\micron\ ``cold core'' map to be 20\%.

\subsection{SCUBA}

Although in previous papers on TMC-1C we used sub-millimeter maps
observed and reduced by us \citep{Schnee05, Schnee07a}, in this paper
we use maps of TMC-1C at 450 and 850~\micron\ from the new SCUBA
Legacy Catalogues \citep{diFrancesco08}.  We choose to use the new
maps because they include all of the data taken with SCUBA
\citep{Holland99} at all epochs and use the most current calibration
and reduction methods.  The noise in the maps are 132 and 21 mJy
beam$^{-1}$ at 450 and 850~\micron, respectively, and have resolutions
of appoximately 9\arcsec\ and 14\arcsec.  The uncertainties in the
absolute flux calibration (primarily due to beam-shape uncertainty and
fluctuations of opacity above the telescope during observations and
calibration), are 20\% at 850~\micron\ and 50\% at 450~\micron, as
described in \citet{diFrancesco08}.  We set the zero point of the
Legacy maps by determining the median flux density in two regions off
the core but nearby and subtracting these values (-0.32 and -0.035 Jy
beam$^{-1}$) from the 450 and 850~\micron\ maps.

\subsection{MAMBO}

We observed the 1200~\micron\ continuum emission from TMC-1C in the
autumns of 2002 and 2003 using the MAMBO-2 array \citep{Kreysa99} on
the IRAM 30~m telescope on Pico Veleta (Spain), producing a map with
10.7\arcsec\ resolution.  Data were taken with on-the-fly mapping,
with the telescope sub-reflector chopping in azimuth by
60\arcsec-70\arcsec, and the resultant image was reconstructed using
the \citet{Emerson79} algorithm iteratively to properly reproduce
large-scale emission.  The noise in map is 3 mJy beam$^{-1}$, and the
flux calibration uncertainty, as derived from the rms of calibrator
observations across pooled observing sessions and the uncertainty in
the intrinsic calibrator fluxes, is about 10\%.  For further details
on the 1200~\micron\ observations and data reduction, see
\citet{Schnee07a} and \citet{Kauffmann08}.

\subsection{Bolocam}

We observed the 2100~\micron\ continuum emission from TMC-1C in the
autumn of 2007 using Bolocam on the Caltech Submillimeter Observatory
(CSO).  Data were collected in raster scan mode at a scan speed of
240\arcsec s$^{-1}$, with a step size between rasters of
11\arcsec. Each raster was 20\arcmin\ in length, and each of the 105
total scans were separated by 11\arcsec.  The resultant map has a
coverage of $\sim$20\arcmin$\times$20\arcmin with 60\arcsec\
resolution.  Three sets of observations contain rasters parallel to
RA, and the other three contain rasters parallel to Dec.  Data
reduction and flux calibration, accomplished through observations of
Uranus and Neptune, was done as described in \citet{Sayers09}.

Observations at approximately 2~mm are dominated by sky noise.
Although the Bolocam pipeline PCA cleaning algorithm efficiently
removes sky noise, it also removes some of the source flux.  This
decreases the peak intensity, introduces negative bowls around bright
sources, and reduces the brightness on scales larger than $1'$ for
extended sources.  We recover this lost flux using a mapping algorithm
that iteratively subtracts a source model from the real data
\citep[for more description and performance evaluation,
see][]{Enoch06}.  This technique is similar to CLEAN \citep{Hogbom74,
Schwarz78}, but works in the image plane, and the real data are used
for the source model rather than, e.g., a Gaussian component model.
The first source model is formed from the original map, including only
regions with signal-to-noise greater than 1.5.  That source model is
subtracted from the original timestream data, which is then re-cleaned
and mapped without the source flux.  This process is iterated until
the source model stabilizes, 10 iterations in this case.  TMC-1C is a
faint source, so the correction is relatively small, with the source
flux density increasing by up to 30\% after iterative mapping.  The
noise is the 2100~\micron\ map is 8~mJy~beam$^{-1}$, and the
uncertainty in the absolute flux calibration is $\sim$20\%.

\section{Analysis} \label{RESULTS}

Here we estimate the temperature, column density and emissivity
spectral index in TMC-1C using our flux density maps at 160, 450, 850,
1200 and 2100~\micron.  Under the assumptions that the dust is
isothermal and characterized by a single emissivity spectral index,
the flux density per beam in each map can be described as a modified
blackbody, given by,

\begin{equation} \label{FLUXEQ}
S_\nu = \Omega B_\nu(T_d) \kappa_\nu \mu m_H N_{H_2}
\end{equation}
where
\begin{equation} \label{PLANCKEQ}
B_\nu(T_d) = \frac{2 h \nu^3}{c^2} \frac{1}{\exp(h \nu / k T_d) - 1}
\end{equation}
and 
\begin{equation} \label{KAPPAEQ}
\kappa_\nu = \kappa_{230} \left(\frac{\nu}{230~{\rm GHz}} \right)^\beta
\end{equation}  

In equation (\ref{FLUXEQ}), $S_\nu$ is the flux density per beam,
$\Omega$ is the solid angle of the beam, $B_\nu(T_d)$ is the blackbody
emission from the dust at temperature $T_d$, $\mu = 2.8$ is the mean
molecular weight of interstellar material in a molecular cloud per
hydrogen molecule, $m_H$ is the mass of the hydrogen atom, $N_{H_2}$
is the column density of hydrogen molecules, and a gas-to-dust ratio
of 100 is assumed.  In equation (\ref{KAPPAEQ}), $\kappa_{230} =
0.009$ cm$^2$ g$^{-1}$ is the emissivity of the dust grains at a gas
density of 10$^6$~cm$^{-3}$ covered by a thin ice mantle at 230~GHz
\citep[][Column 6 of Table 1]{Ossenkopf94} and $\beta$ is the
emissivity spectral index of the dust.  Although Equations
(\ref{FLUXEQ} -- \ref{KAPPAEQ}) are often used to described the
observed emission from starless cores, such observations average over
a range of dust grain properties.  Furthermore, the emissivity
($\kappa_{230}$) is uncertain by roughly a factor of 2.  If we had
chosen a different value for $\kappa_{230}$, this would affect the
derived column density, but would have no impact on the derived
temperature and emissivity spectral index.  We do not consider
variations in $\kappa_{230}$ in this paper, but instead save this
analysis for a future analysis.

\subsection{Color Correction} \label{COLOR}

The fluxes given in the dust emission maps are derived from the total
energy detected convolved with the wavelength-dependent instrumental
response and an assumed SED of the emitting source.  The assumed SEDs
in the {\it Spitzer}, SCUBA, MAMBO and Bolocam maps are that of a
10,000~K blackbody, a flat-spectrum source, a flat-spectrum source and
a blackbody in the Rayleigh-Jeans regime, respectively.  Because we
are observing emission from dust, with an assumed SED of a modified
blackbody, the quoted flux at the nominal wavelength of each map ought
to be corrected.  However, the relatively narrow bandpasses of the
detectors results in a modest correction factor.

For each map, we use the instrumental response and our best-guess dust
SED of a modified blackbody with $T_d = 10$~K and $\beta = 2$ to find
the ``reference'' wavelength of the observations.  The integral of the
convolution of the instrumental response and the dust SED are equal
above and below this reference wavelength.

\begin{equation} \label{REFLAMBDA}
\int_{0}^{\lambda_{\rm ref}} S_{\lambda} R(\lambda) d \lambda = \int_{\lambda_{\rm ref}}^{\infty} S_{\lambda} R(\lambda) d \lambda
\end{equation}
where $S_\lambda$ is the flux emitted by the dust at wavelength
$\lambda$ and $R(\lambda)$ is the instrumental response at wavelength
$\lambda$.

The reference wavelength, used in our calculations of the dust
emission properties, are close to the nominal wavelength of each map,
with the biggest difference in the MAMBO map because of its wide
bandpass.  In all maps, the flux uncertainties are larger than the
correction from using the reference wavelength in our calculations
rather than the nominal wavelength.  The derived reference wavelengths
are given in Table \ref{FLUXTAB}.

\subsection{Single SED} \label{MAP1D}

One way to estimate the dust temperature, column density and
emissivity spectral index is to fit modified blackbody SEDs (given by
Equations \ref{FLUXEQ} -- \ref{KAPPAEQ}) to the five fluxes (at 160,
450, 850, 1200 and 2100~\micron) at each position.  Although the
resolutions of the maps at 450, 850 and 1200~\micron\ are fairly
similar (from 7.5\arcsec\ -- 14\arcsec), the 60\arcsec\ resolution of
the 160 and 2100~\micron\ maps is not a good match to the others.
Here we handle the mismatched beam sizes by making an average SED from
the five flux maps to derive the best-fit dust properties ($T_d$,
$N_{H_2}$ and $\beta$).

\begin{figure}
\epsscale{1.0} 
\plotone{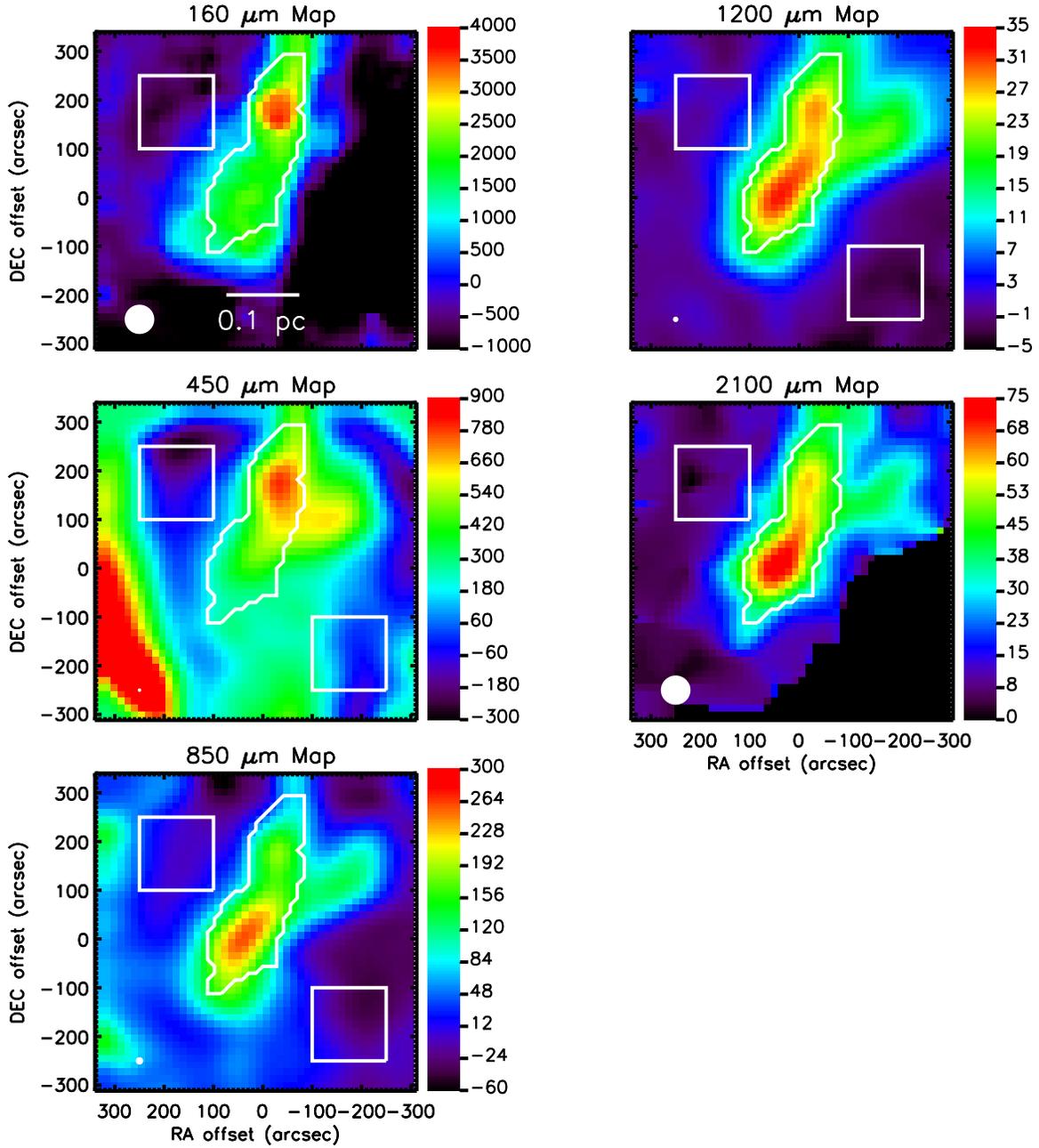}
\caption{Flux maps of TMC-1C at 160, 450, 850, 1200 and 2100~\micron,
in units of mJy beam$^{-1}$, smoothed with the 2100~\micron\ map's
beam.  The white contour shows the region that is above 3$\sigma$ in
the 160, 850, 1200 and 2100~\micron\ maps and above 2$\sigma$ in the
450~\micron\ map, and is the region used for determining the dust
properties.  The white boxes show the regions over which the noise in
the maps was calculated.  The beam size of each map is shown in the
bottom left corners of each panel.
\label{FIVEFLUXPLOT}}
\end{figure}

We first smooth the 450, 850 and 1200~\micron\ maps to the resolution
of the 160 and 2100~\micron\ maps by convolution with a
two-dimensional Gaussian beam.  We then calculate the average flux in
those pixels brighter than 3$\sigma$ in the 160, 850, 1200, and
2100~\micron\ maps, and brighter than 2$\sigma$ in the 450~\micron\
map (which is noisier than the other maps).  The area with significant
emission in all five maps is shown in Figure \ref{FIVEFLUXPLOT}.  The
noise level in the maps is determined in regions off-source but near
TMC-1C (see Figure \ref{FIVEFLUXPLOT}).  We then used a $\chi^2$
minimization routine to fit the five measured fluxes with a modified
blackbody SED, finding the values of $T_d$, $N_{H_2}$ and $\beta$.

\begin{figure}
\epsscale{1.0} 
\plotone{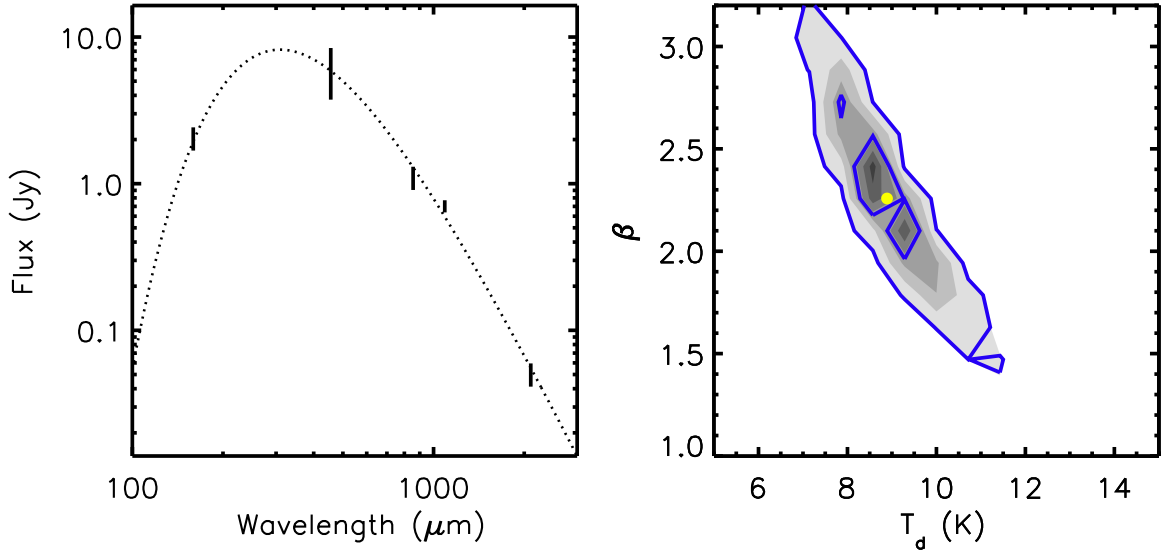}
\caption{({\it left}) The fluxes of TMC-1C at 160, 450, 850, 1200 and
2100~\micron, in a 60\arcsec\ Gaussian beam, averaged over the high
signal-to-noise portion of the maps, shown with 1$\sigma$ error bars.
The dotted line shows the best-fit modified blackbody spectrum, which
has column density $N_{H_2} =$ \SINGLESEDNH, dust temperature $T_d =$
\SINGLESEDTEMP\ and emissivity spectral index $\beta =$
\SINGLESEDBETA. See section \ref{MAP1D} for details.({\it right})
Filled contours show the density of $T_d - \beta$ values derived in
the Monte Carlo simulation of the average TMC-1C fluxes, with the open
blue contours containing 50\% and 95\% of all points.  The yellow
circle shows the position of the median dust temperature and
emissivity spectral index, used to calculate the SED in the left side
of this figure.  The fluxes and uncertainties are given in Table
\ref{FLUXTAB}.
\label{SINGLESEDPLOT}}
\end{figure}

In order to determine how noise and calibration uncertainties affect
the accuracy of the derived dust properties, we run a simple Monte
Carlo simulation.  In each trial, Gaussian random noise of the amount
observed is added to each flux map, and then each map is scaled by a
multiplicative factor $f = 1.0 + \delta$, where $\delta$ is randomly
chosen from a normal distribution of mean zero and standard deviation
equal to the absolute calibration uncertainty in each map.  Each map
is modified by a different scale factor $f$, and this process was
repeated 5000 times.  We find that the best-fit dust properties for
the average SED over TMC-1C are $\beta$ = \SINGLESEDBETA\ $\pm$
\SINGLESEDBETARMS, $T_d$ = \SINGLESEDTEMP\ $\pm$ \SINGLESEDTEMPRMS,
and $N_{H_2}$ = \SINGLESEDNH\ $\pm$ \SINGLESEDNHRMS.  The observed
fluxes and best-fit SED are plotted in Figure \ref{SINGLESEDPLOT} and
the fluxes and uncertainties are given in Table \ref{FLUXTAB}.

A fundamental flaw in this analysis is that equations \ref{FLUXEQ} --
\ref{KAPPAEQ} are only accurate for isothermal dust characterized by a
single emissivity spectral index, but TMC-1C is known to be warmer at
its edges than at the center \citep{Schnee07a}, and it is certainly
possible that $\beta$ also changes within the starless core.  As shown
by \citet{Shetty09a}, the incorrect assumption of isothermal dust
drives an anti-correlation between the derived dust temperature and
emissivity spectral index.  A spurious anti-correlation between $T_d$
and $\beta$ may also arise due to noise in the emission maps
\citep{Shetty09a, Shetty09b}.  An anti-correlation between $\beta$ and
$T_d$ is seen in the Monte Carlo simulation we used to derive the dust
properties, as shown in Figure \ref{SINGLESEDPLOT}.  In the following
section we attempt to mitigate the effects of variations in the dust
properties on our analysis.

\subsection{Maps} \label{MAP2D}

The dust properties ($T_d$, $N_{H_2}$ and $\beta$) in TMC-1C may vary
both in the plane of the sky and along the line of sight.  The
line-of-sight variations in the dust properties cannot be determined
without knowledge of the geometry of the core, but variations in the
plane of the sky are more tractable.  Here we make maps of TMC-1C,
first of the temperature and column density while holding the
emissivity spectral index constant, and then of all three quantities.

\subsubsection{Maps with Constant $\beta$} \label{CONSTANTBETA}

We first regrid the 450, 850 and 1200~\micron\ maps to the resolution
of the 850~\micron\ map, 14\arcsec. We then determine the noise in the
maps in the same regions selected in Section \ref{MAP1D} and calculate
the dust temperature and column density in those pixels brighter than
3$\sigma$ in the 850 and 1200~\micron\ maps and brighter than
2$\sigma$ in the 450~\micron\ map, using equations \ref{FLUXEQ} --
\ref{KAPPAEQ} for an assumed value of the emissivity spectral index,
as was done in \citet{Schnee07a}.  Temperature and column density maps
are made for each value of the emissivity spectral index in the range
$1.0 \le \beta \le 4.0$ in steps of 0.1. To account for the effects of
noise and calibration uncertainties, we use the same Monte Carlo
technique described in Section \ref{MAP1D} to add Gaussian random
noise to each pixel and adjust each map by a multiplicative scale
factor, and this process is repeated 100 times for each value of
$\beta$.

\begin{figure}
\epsscale{0.8} 
\plotone{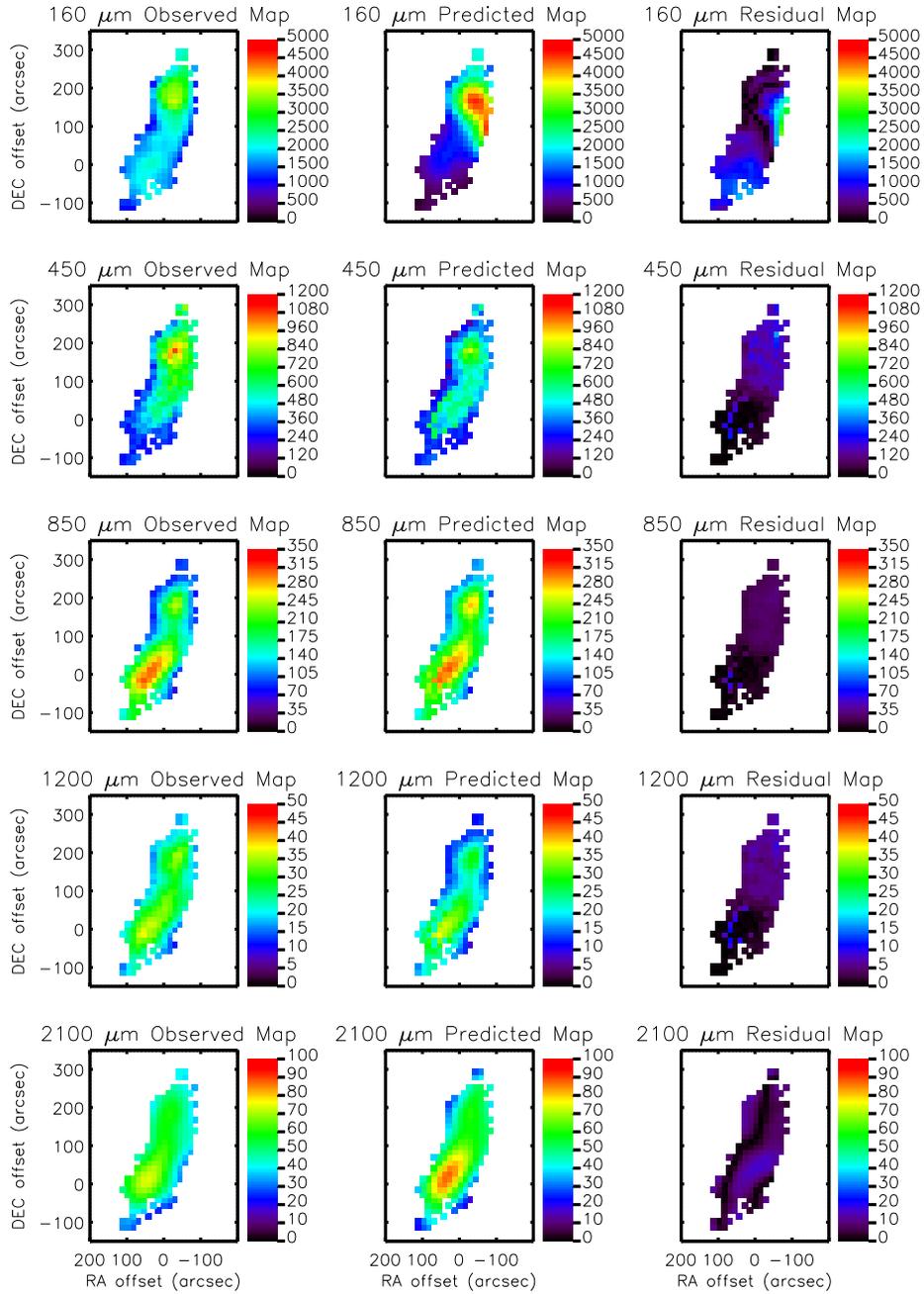}
\caption{({\it left column}) The observed maps of TMC-1C at 160, 450,
850, 1200 and 2100 \micron, shown in units of mJy beam$^{-1}$.  ({\it
center column}) The predicted maps of TMC-1C, based on the derived
temperature and column density maps (see Figure
\ref{MAPTDAVBETAPLOT}), with $\beta = $\MAPSINGLEBETA.  ({\it right
column}) The absolute value of the residuals.  See section
\ref{CONSTANTBETA} for details.
\label{MAP2DPLOT}}
\end{figure}

To determine which value of $\beta$ is most appropriate for TMC-1C, we
then use the derived $T_d$ and $N_H$ maps to predict the fluxes at
160, 450, 850, 1200 and 2100~\micron.  The predicted 160 and
2100~\micron\ fluxes are convolved with a 60\arcsec\ Gaussian beam to
match the resolution of the observed maps.  We find that the 160 and
2100~\micron\ maps are best reproduced by an emissivity spectral index
of $\beta =$ \MAPSINGLEBETA\ $\pm$ \MAPSINGLEBETARMS.  The observed,
predicted and residual flux maps for $\beta =$ \MAPSINGLEBETA\ are
shown in Figure \ref{MAP2DPLOT}, and the derived temperature and
column density maps are shown in Figure \ref{MAPTDAVBETAPLOT}.  The
column density is expressed in units of $A_V$, using the conversion
factor $N_{H_2} = 9.4 \times 10^{20}$ cm$^{-2}$ ($A_V$/mag)
\citep{Bohlin78}.

\subsubsection{Maps with Variable $\beta$} \label{VARIABLEBETA}

In section \ref{CONSTANTBETA} we assumed that $\beta$ is constant
across TMC-1C to make 14\arcsec\ maps of the dust temperature and
column density in TMC-1C.  Here we make two-dimensional maps of the
dust temperature, column density {\it and} emissivity spectral index
in TMC-1C, with the tradeoff that the resultant maps have coarser
resolution (60\arcsec).

We first regrid the 450, 850 and 1200~\micron\ maps to the 60\arcsec\
resolution of the 160 and 2100~\micron\ maps.  For each pixel brighter
than 3$\sigma$, we then calculate $T_d$, $N_{H_2}$ and $\beta$ using
equations \ref{FLUXEQ} -- \ref{KAPPAEQ}.  To estimate the
uncertainties in this method due to noise and absolute calibration
errors, we used the same Monte Carlo techniques as in sections
\ref{MAP1D} and \ref{CONSTANTBETA} with 1000 trials.  In each trial
there were about 15 pixels with sufficiently high signal to noise to
calculate the dust properties, and within the maps there is a clear
anti-correlation between $T_d$ and $\beta$, as shown in Figure
\ref{MAPTDAVBETAPLOT}.  The median emissivity spectral index and dust
temperature derived from the high signal to noise points in all trials
are $\beta =$ \MAPMULTIBETA\ $\pm$ \MAPMULTIBETARMS\ and $T_d =$
\MAPMULTITEMP\ $\pm$ \MAPMULTITEMPRMS, similar to that derived from
the composite SED of TMC-1C in section \ref{MAP1D}.

\section{Discussion} \label{DISCUSSION}

In Section \ref{VARIABLEBETA} we found that there is an
anti-correlation between the dust temperature and emissivity spectral
index in the two dimensional maps of TMC-1C.  Although this phenomenon
has been reported in other star-forming regions \citep{Dupac03} and in
other galaxies \citep{Yang07}, this is not a known feature of starless
cores.  The assumption that the temperature along each line-of-sight
is constant, even though the temperature profile in TMC-1C is very
likely to be colder at higher densities, can lead to the mistaken
conclusion that there is an anti-correlation between $T_d$ and $\beta$
\citep{Shetty09a, Shetty09b}.  Noise in the dust emission maps can
also drive a spurious $T_d - \beta$ anti-correlation.  Here we
investigate whether the dust temperature and emissivity spectral index
are truely anti-correlated, and conclude that this is not necessarily
the case.  We then determine what constant value of $\beta$ best
characterizes TMC-1C.

\subsection{Testing the $T_d - \beta$ Anti-correlation}

To determine the effects of temperature variations along the line of
sight and noise on the dust properties derived from flux maps at 160,
450, 850, 1200 and 2100~\micron, we consider a spherical model of a
starless core with plausible properties.  The density in the core is
given by
\begin{equation} \label{DENSITYEQ}
n(r) = \frac{n_0}{1 + (r/r_0)^{2.5}}
\end{equation}
where $r$ is the radius, $n_0 = 10^6$~cm$^{-3}$ and $r_0 = 0.005$~pc.
The temperature in the core is given by
\begin{equation} \label{TEMPEQ}
T_d(r) = T_0 \left(1 + r/r_1 \right)
\end{equation}
where $T_0 = 6.4$~K and $r_1 = 0.15$~pc.  The emissivity spectral
index is constant in the core at a value $\beta = 2.2$.  This model
core is colder and denser at the center, as expected for an externally
heated starless core.  We do not claim that these are the temperature
and density profiles for TMC-1C, but there is a general agreement
between the shape of the model SED and that of TMC-1C.  The density
and temperature profiles are shown in Figure \ref{MODELSPHEREPLOT}, as
is the total emergent SED.

For each line of sight passing through the model core, we determine
the volume of intersection between a cylinder and each spherical shell
of the core.  Given the density and temperature distributions from
Equations \ref{DENSITYEQ} and \ref{TEMPEQ}, the flux from each layer
is given by
\begin{equation} \label{SHELLEQ}
S_\nu = \frac{B_\nu(T_d) \kappa_\nu n_r V_r}{d^2}
\end{equation}
where $V_r$ is the volume of intersection and $d$ is the distance to
the core.  The two-dimensional flux distribution is calculated by
summing the flux from each layer along each line of sight.  For the
sake of convenience, the total flux density of the model core, summed
over all radii, is scaled by a common factor such that the peak of the
SED has a value of unity.

We ``observe'' the model core by deriving two-dimensional flux maps at
160, 450, 850, 1200 and 2100~\micron\ and adding Gaussian random noise
equal to 5\% of the peak flux in each map.  Using equations
\ref{FLUXEQ} -- \ref{KAPPAEQ}, we then derive two-dimensional maps of
the column density, dust temperature and emissivity spectral index,
assuming that these quantities are constant along the line of sight
(as was done in Section \ref{VARIABLEBETA}).  The derived column
density map peaks at the center of the core, where the temperature is
seen to decrease, as one would expect from the input density and
temperature profiles (see Figure \ref{MODELSPHEREPLOT}).  The derived
emissivity spectral index map covers a wide range of values, mostly
between $1.0 \le \beta \le 3.5$, even though the input emissivity
spectral index was held constant.  The median derived emissivity
spectral index is \MODELBETA\ $\pm$ \MODELBETARMS, which is close to
the input value of $\beta = 2.2$.

Furthermore, the derived temperature and emissivity spectral index are
seen to be anti-correlated.  The combination of noise and temperature
variations along the line of sight have been shown to drive an
anti-correlation between these two quantities \citep{Shetty09a,
Shetty09b}, so it is not surprising that this effect is seen here.
Interestingly, the portion of the $T_d - \beta$ parameter space
covered by the model starless core overlaps significantly with that of
TMC-1C (see Figures \ref{VARIABLEBETA} and \ref{MODELSPHEREPLOT}).
From this we conclude that there is no definitive evidence that the
emissivity spectral index in TMC-1C varies with temperature, and that
the more reliable estimates of column density and temperature will
come from fitting the observed fluxes while holding the emissivity
spectral index constant.

\subsection{Average Emissivity Spectral Index in TMC-1C}

Estimates of the emissivity spectral index in TMC-1C derived from its
average SED (see Section \ref{MAP1D}) and from two dimensional maps
using the 450, 850 and 1200~\micron\ maps to predict the lower
resolution 160 and 2100~\micron\ maps (see Section
\ref{CONSTANTBETA}), $\beta = \SINGLESEDBETA \pm \SINGLESEDBETARMS$
and $\beta = \MAPSINGLEBETA \pm \MAPSINGLEBETARMS$, are in good
agreement with each other.  Given that an analysis of the flux maps
from an idealized core, with line-of-sight temperature variations and
noise, gives the correct {\it average} emissivity spectral index
(within errors), we suggest that the value of the emissivity spectral
index in TMC-1C is well-represented by $\beta \simeq 2.2 \pm 0.5$.
This is consistent with the emissivity spectral index in the diffuse
ISM of $\beta = 2$ \citep{Draine07, Draine84} and with the emissivity
spectral index measured in the starless core L1498 \citep[$\beta =
2.44 \pm 0.62$,][]{Shirley05}.  However, due to the large uncertainty
in $\beta$ in this study, we cannot rule out the possibility that the
emissivity spectral index in TMC-1C could also be intermediate between
the shallower SEDs seen in circumstellar disks and the SED of the ISM.

When deriving the dust temperature and mass of prestellar cores from
sparsely-sampled SEDs, it is common to assume that the emissivity
spectral index of the dust is equal to 2 \citep[e.g.][]{Kirk07,
Simpson08}.  Observations of L1498 \citep{Shirley05} and our
observations of TMC-1C are consistent with $\beta = 2$, though the
best-fit values of the emissivity spectral index in both cores are
slightly steeper, which would result in lower temperatures and higher
masses.  However, with so few measurements of $\beta$ in starless
cores, there is not sufficient information with which to re-examine
the dust properties in previously published studies.  Upcoming surveys
of the Gould Belt molecular clouds using {\it
Herschel}\footnote[1]{http://starformation-herschel.iap.fr/gouldbelt}
and {\it
SCUBA-2}\footnote[2]{http://www.jach.hawaii.edu/JCMT/surveys/gb/}
promise to make statistically significant measurements of the
emissivity spectral index in starless cores.

\section{Summary} \label{SUMMARY}

In this paper we have analyzed the dust emission at 160, 450, 850,
1200 and 2100~\micron\ from TMC-1C.  Our previous studies of this
starless core have mapped the dust temperature and column density.
Here, using new 160 and 2100~\micron\ maps and an improved reduction
of the 450 and 850~\micron\ maps, we are able to estimate the
emissivity spectral index and make an improved analysis of the
temperature and column density in TMC-1C.

We find that the spectral index of the dust emission, as calculated by
a variety of related methods, is in the range $1.7 \le \beta \le 2.7$
This is consistent with the value of the emissivity spectral index
often assumed in studies of starless cores, and is also consistent
with laboratory measurements and observations of the behavior of dust
grains at low temperatures ($T_d \le 10$~K).  However, there is still
a relatively large uncertainty in the emissivity spectral index.

\begin{figure}
\epsscale{0.8} 
\plotone{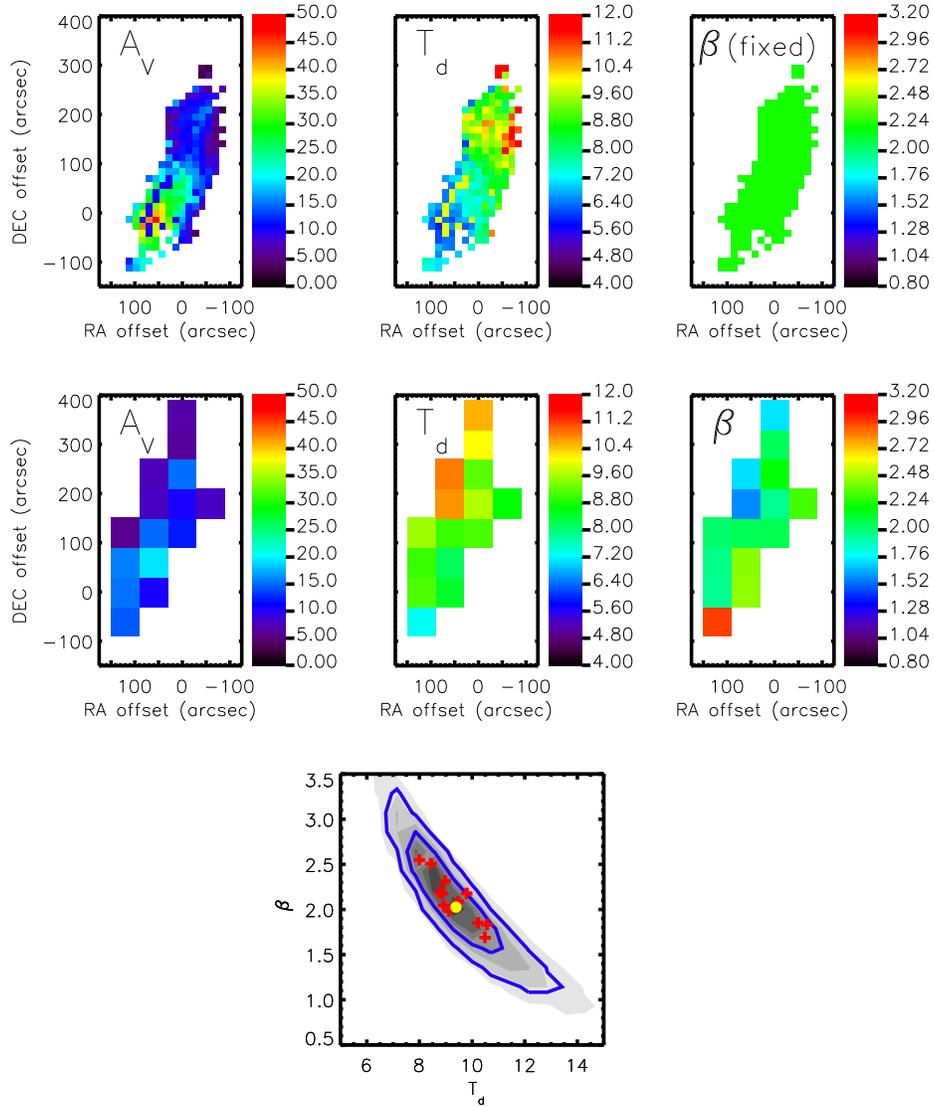}
\caption{({\it top row}) Column density and dust temperature maps,
expressed in units of $A_V$ and K, derived from 14\arcsec\ maps of
TMC-1C at 450, 850 and 1200~\micron\ using $\beta =
$\MAPSINGLEBETA. See section \ref{CONSTANTBETA} for details. ({\it
center row}) Column density, dust temperature and emissivity spectral
index maps of TMC-1C, derived from 60\arcsec\ maps at 160, 450, 850,
1200 and 2100~\micron.  See section \ref{VARIABLEBETA} for details.
({\it bottom row}) Filled contours show the density of $T_d - \beta$
values, with the open blue contours containing 50\% and 95\% of all
points.  The yellow circle shows the position of the median dust
temperature ($T_d =$ \MAPMULTITEMP) and emissivity spectral index
($\beta =$ \MAPMULTIBETA).  The red crosses show the values of $\beta$
and $T_d$ derived from each independent position in the 60\arcsec\
flux density maps with high signal to noise.
\label{MAPTDAVBETAPLOT}}
\end{figure}

\begin{figure}
\epsscale{1.0} 
\plotone{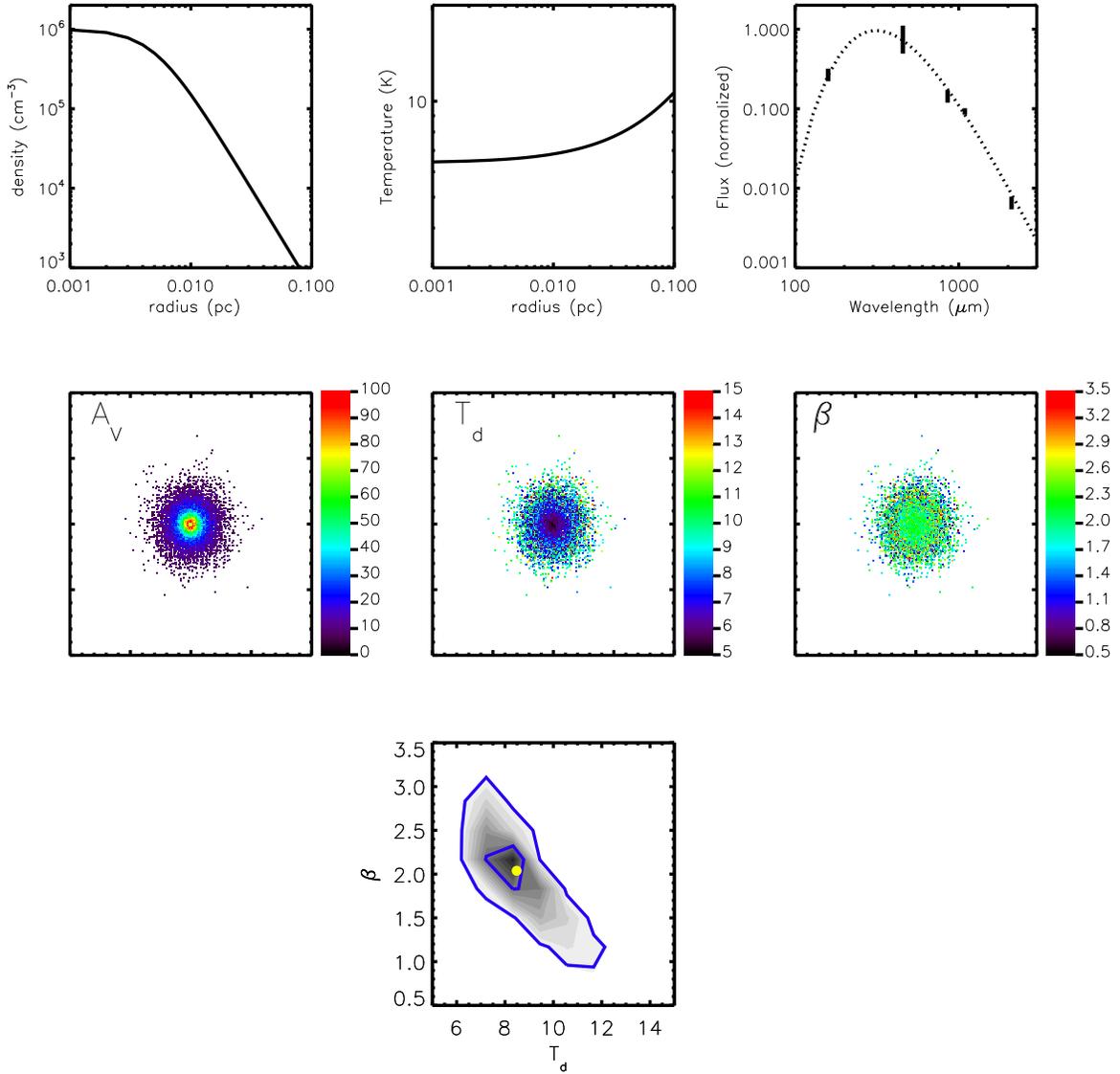}
\caption{({\it top left and center}) Density and temperature profiles
of a spherical model of a starless core.  ({\it top right}) The SED of
the model starless core with $\beta = 2.2$, normalized such that the
SED peak is at unity.  The TMC-1C observations are also plotted for
reference, and scaled by a common factor such that the 450~\micron\
flux from TMC-1C is equal to that of the model core.  ({\it center
row}) Column density, temperature and emissivity spectral index maps
that would be derived from the noisy maps of the model starless core.
({\it bottom row}) Filled contours show the density of $T_d - \beta$
values, with the open blue contours containing 50\% and 95\% of all
points.  The yellow circle shows the position of the median dust
temperature and emissivity spectral index.  The anti-correlation
between $T_d$ and $\beta$ is similar to that observed in TMC-1C (see
Figure \ref{MAPTDAVBETAPLOT}).
\label{MODELSPHEREPLOT}}
\end{figure}

Using the new estimate of the emissivity spectral index in TMC-1C
($\beta = $\MAPSINGLEBETA), we construct temperature and column
density maps, which reach values of $T_d \simeq 7~K$ and $A_V \simeq
50$ at the position of the dust peak, as shown in Figure
\ref{MAPTDAVBETAPLOT}.  We find no conclusive evidence that $\beta$
varies within the core, though line-of-sight temperature variations
and noise in the flux maps leave open the possibility that the
emissivity spectral index is anti-correlated with dust temperature.

\acknowledgments

We thank our referee, Yancy Shirley, for comments that improved the
clarity of this paper.  SS acknowledges support from the Owens Valley
Radio Observatory, which is supported by the National Science
Foundation through grant AST 05-40399.  JS was partially supported by
a NASA Postdoctoral Program Fellowship.  Support was provided to ME by
NASA through the {\it Spitzer Space Telescope} Fellowship Program.
The JCMT is operated by the Joint Astronomy Centre on behalf of the
Particle Physics and Astronomy Research Council of the United Kingdom,
the Netherlands Organisation for Scientific Research, and the National
Research Council of Canada.  IRAM is supported by INSU/CNRS (France),
MPG (Germany), and IGN (Spain).  The CSO is supported by the NSF fund
under contract AST 02-29008.

{}

\begin{deluxetable}{cccc} 
\tablewidth{0pt}
\tabletypesize{\scriptsize}
\tablecaption{TMC-1C Fluxes \label{FLUXTAB}}
\tablehead{
 \colhead{Nominal Wavelength}   & 
 \colhead{Reference Wavelength} & 
 \colhead{Average Flux}         & 
 \colhead{Flux Uncertainty}     \\
 \colhead{(\micron)}            &
 \colhead{(\micron)}            &
 \colhead{(Jy/beam\tablenotemark{1})} &
 \colhead{}}
\startdata
 160 &  159 &  2.0   & 20\% \\
 450 &  456 &  5.6   & 50\% \\
 850 &  856 &  1.1   & 20\% \\
1200 & 1090 &  0.70  & 10\% \\
2100 & 2103 &  0.050 & 20\% \\
\enddata
\tablenotetext{1}{in a 60\arcsec\ Gaussian beam}
\end{deluxetable}

\end{document}